\documentclass{MITcsail} 

\usepackage{hyperref}
\usepackage{url}

\usepackage{amsmath,amsfonts,bm}

\def\eqref#1{equation~\ref{#1}}

\def\1{\bm{1}}

\DeclareMathAlphabet{\mathsfit}{\encodingdefault}{\sfdefault}{m}{sl}
\SetMathAlphabet{\mathsfit}{bold}{\encodingdefault}{\sfdefault}{bx}{n}

\definecolor{deepred}{rgb}{0.631,0.102,0.102}
\definecolor{amethyst}{rgb}{0.6, 0.4, 0.8}
\definecolor{darkgreen}{rgb}{0.3,0.7,0.3}
\definecolor{salmon}{RGB}{241, 150, 141}
\definecolor{mildyellow}{HTML}{FFF2CC}
\newcommand{\finding}[2]{
\vspace{12pt}
\noindent\fcolorbox{deepred}{mildyellow}{\begin{minipage}{0.98\columnwidth}
    \textcolor{deepred}{\textbf{\textit{Finding} #1.} #2}
\end{minipage}}
\vspace{6pt}
}

\usepackage{amsmath, amssymb, amsfonts}
\usepackage{listings}
\usepackage{xcolor}
\usepackage{geometry}
\lstdefinestyle{cpp}{
  language=C++,
  basicstyle=\ttfamily\footnotesize,
  keywordstyle=\color{blue},
  commentstyle=\color{gray},
  stringstyle=\color{red},
  showstringspaces=false,
  numbers=left,
  numberstyle=\tiny\color{gray},
  breaklines=true,
  frame=single,
  tabsize=2
}

\definecolor{darkred}{RGB}{140, 21, 21}
\definecolor{lightgray}{gray}{0.7}
\definecolor{orange}{HTML}{F58025}

\usepackage{hyperref}
\usepackage{xcolor}
\hypersetup{
    colorlinks=true,
    linkcolor=orange,
    citecolor=lightgray,
    filecolor=darkred,
    urlcolor=orange
}
\usepackage[numbers, sort&compress]{natbib}
\usepackage{amsmath}
\usepackage{amsfonts}
\usepackage{amssymb}
\usepackage{wrapfig}
\usepackage{subcaption}
\usepackage{multirow}
 \usepackage{mathtools} 

\usepackage{verbatim}

\usepackage{anyfontsize}

\usepackage{microtype}
\usepackage{graphicx}
\usepackage{booktabs} 
\usepackage{multirow}
\usepackage{amsmath,amssymb}
\usepackage{booktabs}
\usepackage{caption,subcaption}
\usepackage{svg}
\usepackage{authblk}

\usepackage[ruled,vlined,linesnumbered]{algorithm2e}
\SetKwComment{Comment}{$\triangleright$ }{}
\SetKwInput{KwInput}{Input}
\SetKwInput{KwOutput}{Output}

\definecolor{mygreen}{HTML}{3cb44b}
\definecolor{skyblue}{HTML}{beffff}
\definecolor{lightgreen}{HTML}{90ee90}

\usepackage{color, colortbl}

\definecolor{emerald}{rgb}{0.31, 0.78, 0.37}

\usepackage{datetime}
\newdateformat{ymd}{\the\year-\the\month-\the\day}

\usepackage{tcolorbox}
\usepackage{enumitem}
\definecolor{mygreen}{HTML}{3cb44b}
\colorlet{myyellow}{green!10!orange!90!}
\makeatletter

\usepackage{tikz}
\usetikzlibrary{arrows,shapes,snakes,automata,backgrounds,fit,petri}
\usepackage{adjustbox}

\newcommand{\RN}[1]{%
	\textup{\lowercase\expandafter{\it \romannumeral#1}}%
}
\usepackage{tabu}

\newcommand{\beq}{\vspace{0mm}\begin{equation}}
\newcommand{\eeq}{\vspace{0mm}\end{equation}}
\newcommand{\beqs}{\vspace{0mm}\begin{eqnarray}}
\newcommand{\eeqs}{\vspace{0mm}\end{eqnarray}}
\newcommand{\barr}{\begin{array}}
\newcommand{\earr}{\end{array}}

\usepackage{color, colortbl}
\definecolor{Gray}{gray}{0.93}

\usepackage{lipsum}

\usepackage{pifont}

\usepackage{makecell}

\usepackage{xcolor,amsmath}
\usepackage[linesnumbered,ruled,vlined]{algorithm2e}
\DontPrintSemicolon

\usepackage{xcolor}
\definecolor{mygreen}{HTML}{3cb44b}

\SetKwComment{Comment}{\color{green!50!black}\# }{}

\SetKwProg{Function}{def}{:}{}

\SetKwProg{For}{for}{:}{}
\SetKwProg{If}{if}{:}{}

\usepackage{epigraph}
\usepackage{xspace}

\newcommand{\model}{AutoCode\xspace} 

\newcommand{\pagehead}{\model}

\title{AutoCode: LLMs as Problem Setters for \\[0.3em]
Competitive Programming}

\author[1,$*$]{Shang Zhou}
\author[2,$*$]{Zihan Zheng}
\author[3,$*$]{Kaiyuan Liu}
\author[4,$*$]{Zeyu Shen}
\author[4,$*$]{Zerui Cheng}
\author[1]{Zexing Chen}
\author[5]{Hansen He}
\author[4]{\mbox{Jianzhu Yao}}
\author[7]{Huanzhi Mao}
\author[7]{Qiuyang Mang}
\author[6]{Tianfu Fu}
\author[8]{Beichen Li}
\author[9]{Dongruixuan Li}
\author[4,$\dagger$]{\\Wenhao Chai}
\author[4,$\dagger$]{\mbox{Zhuang Liu}}
\author[4,$\dagger$]{Aleksandra Korolova}
\author[4,$\dagger$]{\mbox{Peter Henderson}}
\author[3,$\dagger$]{Natasha Jaques}
\author[4, 10, $\dagger$]{\mbox{Pramod Viswanath}}
\author[2,$\dagger$]{\mbox{Saining Xie}}
\author[1,$\dagger$]{Jingbo Shang}

\affil[1]{University of California San Diego}
\affil[2]{New York University}
\affil[3]{University of Washington}
\affil[4]{Princeton University}
\affil[5]{Canyon Crest Academy}
\affil[6]{OpenAI}
\affil[7]{University of California Berkeley}
\affil[8]{Massachusetts Institute of Technology}
\affil[9]{University of Waterloo}
\affil[10]{Sentient Labs}

\newcommand{\correspondence}{
  {
   \raggedright
   \normalfont\fontsize{8}{10}\selectfont
   \textbf{Correspondence:} \{shz060, jshang\}@ucsd.edu \par}
}

\newcommand{\tags}{
  {\vspace{0.5em}
   \raggedright
   \normalfont\fontsize{8}{10}\selectfont
   $^*$Equal Contributions, $^\dagger$Advisors 
   \par}
}

\begin{document}

\maketitle
\correspondence
\tags

\thispagestyle{firstpagestyle} 

\begin{abstract}
\textbf{Abstract:} Writing competitive programming problems is exacting. Authors must: set constraints, input distributions, and edge cases that rule out shortcuts; target specific algorithms (e.g., max-flow, dynamic programming, data structures); and calibrate complexity beyond the reach of most competitors. We argue that this makes for an ideal test of general large language model capabilities and study whether they can do this reliably. We introduce \model, which uses multiple rounds of validation to yield competition-grade problem statements and test cases. On held-out problems, \model test suites approach 99\% consistency with official judgments, a significant improvement over current state-of-the-art methods like HardTests, which achieve less than 81\%. Furthermore, starting with a random seed problem, \model can create novel variants with reference and brute-force solutions. By cross-verifying these generated solutions against test cases, we can further filter out malformed problems. 
Our system ensures high correctness, as verified by human experts.
\model successfully produces novel problems judged by Grandmaster-level (top 0.3\%) competitive programmers to be of contest quality.

\vspace{2mm}
\textbf{Project page:} \href{https://livecodebenchpro.com/projects/autocode/overview}{https://livecodebenchpro.com/projects/autocode/overview}

\end{abstract}

\section{Introduction}

\epigraph{\mbox{An AI system can create and maintain knowledge only to the extent that it can verify that knowledge itself.}}{--- Rich Sutton, \textit{Verification, The Key to AI}, 2001}

As Einstein and Infeld put it, ``The formulation of a problem is often more essential than its solution, which may be merely a matter of mathematical or experimental skill. To raise new questions, new possibilities, to regard old problems from a new angle requires creative imagination and marks real advances in science.''~\citep{EinsteinInfeld1938}.
As large language models (LLMs) march toward general-purpose capabilities, with the ultimate goal of artificial general intelligence (AGI), we argue that testing \textit{problem generation} abilities is just as important as \textit{problem solving} abilities. This is particularly true when applying LLMs to advanced programming tasks, where future advancement and economic integration of LLM coding capabilities will require significant validation.

First, problem setting for competitive coding requires a deeper understanding of algorithms that problem solving may not.
For example, basic problems can collapse into recognizable templates that can be solved with simple tricks; and many standard programming questions often allow for partial credit or boilerplate solutions that can mask incorrect reasoning.
Competitive programming problems have a strict bar designed to assess a deeper understanding of the underlying algorithm design principles, data structures, and complexity trade-offs.
Verifying the vast space of possible solutions, along with sufficient coverage of short-cuts or corner cases is challenging, but a necessity for competition programming problems. As such, problem setting encompasses all the challenges of solving a problem, and then more.

Second, better problem setting will lead to more rigorous competitive programming benchmarks. Since official test data from premier platforms such as Codeforces and AtCoder are not publicly available, researchers currently rely on synthesized datasets such as CodeContests+~\cite{wang2025codecontests+}, TACO~\cite{li2023taco}, and HardTests~\cite{he2025hardtests}. Our analysis (\S \ref{sec:bench}), however, shows that existing test datasets can have both high false positive rates (FPR) and false negative rates (FNR). For instance, a greedy algorithm with poor time complexity might pass a suite of small, random tests, only to fail against adversarially constructed cases designed to expose its flaws. 
This critical weakness creates a distorted evaluation landscape, rewarding models that discover shortcuts.

\begin{figure}[t]
    \centering
    \includegraphics[width=\linewidth]{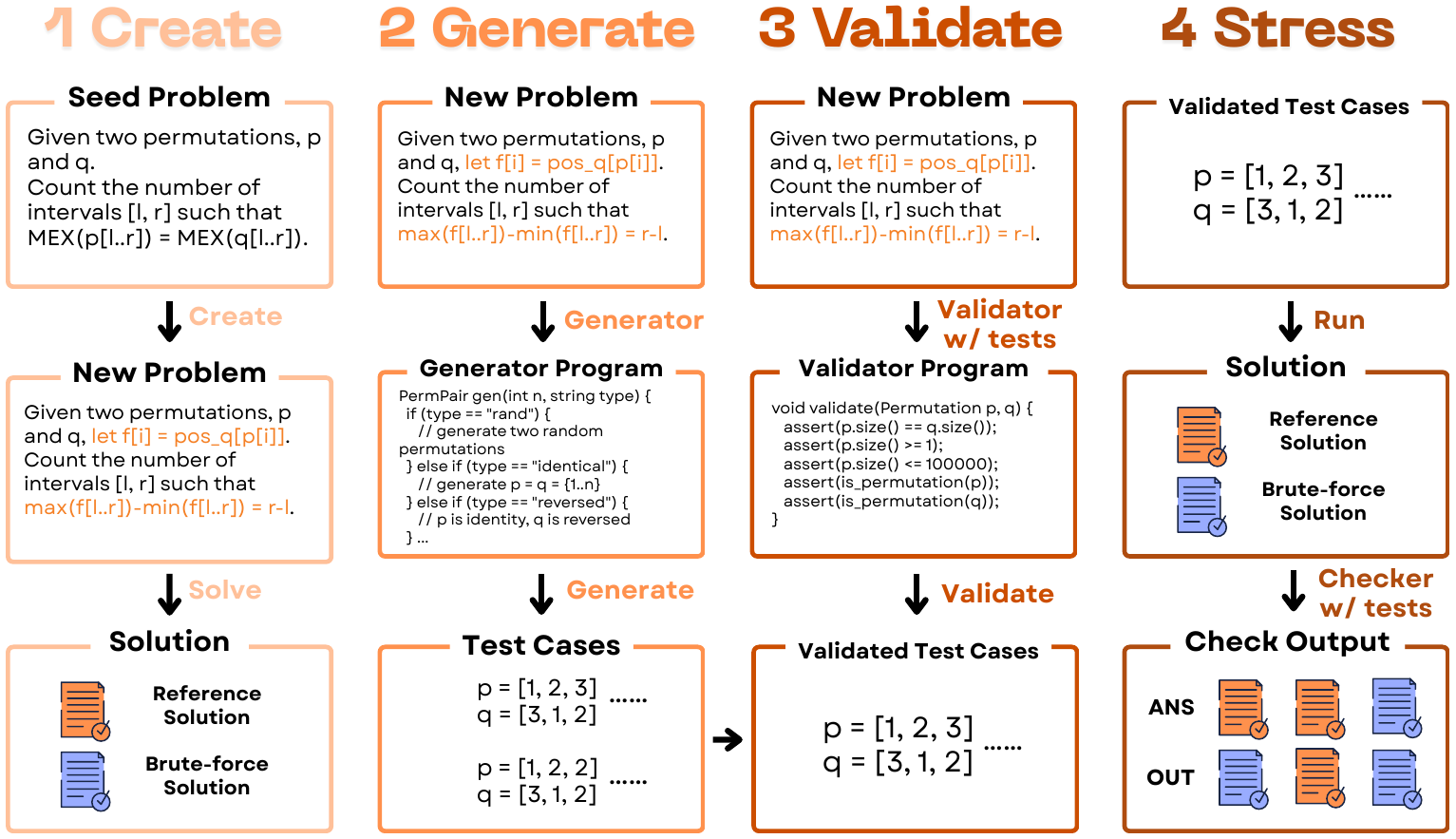}
    \caption{\textbf{\model} introduces a closed-loop multi-role Validator-Generator-Checker framework that enables robust test case generation and scalable, self-verified problem generation for competitive programming. It achieves 98.7\% consistency with official judgments. This framework precisely mirrors the process human experts follow when creating programming contest problems.}
    \label{fig:teasor}
    \vspace{-12pt}
\end{figure}

Third, successful problem setting of novel challenges may pave the path for self-improving models and AGI, as well as validating deployments in complex software stacks. 

LLMs are already used by over a billion people and increasingly for coding applications, saving hundreds of millions of dollars with LLM-assisted programming \citep{openai2025usage,techcrunch2025copilot,aws2024qmilestone}, albeit largely on relatively simple tasks. To expand LLM capabilities toward the strategic reasoning and long-term planning that real programming demands, and to integrate LLMs safely within the software development stack, we must be able to determine, with high fidelity, whether model-generated code is a valid solution to the problem at hand. Misspecified rewards, in the form of malformed problem formulation or verification, can cause models to optimize for the wrong thing or fall into bad local optima. We find, for example, that current benchmark datasets suffer from an even higher FNR than FPR, often caused by invalid inputs to unit tests; as a result, a perfectly valid, creative solution can be unjustly penalized when it crashes or produces an incorrect output on malformed input data. This pollutes data for reinforcement learning by punishing valid lines of reasoning, while high FPR fails to penalize flawed ones.

To address these critical gaps, we introduce \model, a systematic framework that employs LLMs in a closed-loop, multi-role system to automate the entire lifecycle of competitive programming problem creation and evaluation. Our first contribution is an enhanced Validator-Generator-Checker framework that achieves state-of-the-art reliability in test case generation, where the Generator generates test cases, the Validator validates whether the generated test cases satisfy the problem constraints, and the Checker checks whether a solution is correct given the test cases. We move beyond standard implementations with several key heuristics: the LLM generates multiple candidate Validators and Checkers, and we select the most robust one through targeted testing. The generator executes a multi-pronged strategy, creating a diverse and adversarial test suite that includes small-data exhaustion, randomized stress tests, and TLE-inducing cases designed to expose incorrect time complexities. Building on this highly reliable verification pipeline, we introduce our second major contribution: a novel process for generating new, high-quality problems. This process begins with a seed problem to inspire the LLM in a promising direction. To ensure correctness without human intervention, we employ a dual verification protocol: the LLM generates a problem statement, an efficient reference solution, and a simple brute-force solution. The new problem is accepted only after rigorously verifying that the efficient solution's output matches the ground truth established by the brute-force solution across all test cases. In summary, our work makes the following contributions: 
\begin{itemize}[leftmargin=9pt]

\item \textbf{State-of-the-Art Test Cases Generation for Existing Problems.} Our testcase generation framework achieves over 91\% consistency with official judgments on a large-scale benchmark of 7538 problems, significantly outperforming existing methods whose consistency ranges from 72 -- 81\%. We also point out that the ability to generate test cases is not only useful for competitive programming, but also has broad practical significance for general tasks involving input-output matching.

\item \textbf{Novel and High-Quality Problems Generation.} We pioneer a systematic process that uses a dual-verification mechanism to generate, validate, and score new problems. Vetted by elite competitive programmers, this process produces novel problems deemed high-quality and original enough for official contests, with problems passing automated verification achieving a correctness rate of 94\%.

\end{itemize}

\section{Related Works}
LLMs show rapid progress in code generation recently~\cite{hui2024qwen2,deepseek-coder,zhu2024deepseek,gong2025diffucoder,xie2025dream}. Benchmarks and tooling for code reasoning span three threads. Datasets like LiveCodeBench~\cite{jain2024livecodebench} evaluate coding interview problems, LiveCodeBench Pro~\cite{zheng2025livecodebench} focuses on competitive programming, FormulaOne~\cite{beniamini2025formulaone} assesses challenging domain-specific problems, and ELABORATION~\cite{yang2025elaboration} explores human-LLM collaboration. Second, to strengthen verdict reliability, EvalPlus~\cite{liu2023your} augments unit tests, and TACO~\cite{li2023taco}, CodeContests+~\cite{wang2025codecontests+}, HardTests~\cite{he2025hardtests}, TestCase-Eval~\cite{cao2025can}, LogiCase~\cite{sung2025logicase} and the meta-benchmark TCGBench~\cite{ma2025rethinking} propose rule or LLM-driven generators emphasizing constraint validity, edge-case coverage, and adversariality. Third, solver/data-centric lines AlphaCode~\cite{li2022competition}, AceReason~\cite{chen2025acereason}, Absolute Zero~\cite{zhao2025absolute}, and rStar-Coder~\cite{liu2025rstar} scale solution search or curate verified corpora via RL/self-play signals~\cite{zelikman2022starbootstrappingreasoningreasoning,zelikman2024quietstarlanguagemodelsteach,shinn2023reflexionlanguageagentsverbal,madaan2023selfrefineiterativerefinementselffeedback,chen2025spcevolvingselfplaycritic} rather than offering end-to-end test pipelines~\cite{zhang2023algo}. In contrast, \model~unifies and extends these directions by coupling a Validator-Generator-Checker (and interactor) loop that enforces legality plus adversarial coverage with a dual-verification protocol (reference vs. brute force) to generate and certify new problems, addressing both static-benchmark contamination and under-constrained tests within a single framework.
\section{Test Case Generation}
\label{sec:test_case_gen}

Our test case generation process is a structured framework designed for maximum rigor and coverage. The framework as shown in Figure~\ref{fig:teasor} starts with the Validator, which serves as the cornerstone of the entire system. Its function is to ensure any given input strictly adheres to all constraints specified in the problem description. A Validator is critical for minimizing the FNR, as it prevents correct programs from failing on malformed data. Next, the Generator employs diverse strategies to create a broad spectrum of inputs, aiming to reduce the FPR, where incorrect or inefficient programs are erroneously judged as correct. Any invalid cases produced by the Generator are subsequently filtered by the Validator, ensuring we obtain a high-quality set of inputs. Finally, to assess the contestant's output, a Checker compares it against the reference solution's output, while for interactive tasks, an Interactor engages in a multi-turn dialogue with the contestant's program to issue a final verdict. Terminology is defined in Appendix~\ref{preliminaries}. Because one of our prominent goals is to serve as a high-quality verifier for RLVR, we pay particular attention to reducing the FPR. We distinguish test cases (input-answer pairs) from the test data, which includes the Checker and Interactor programs required for evaluation.

\subsection{Validator}

\begin{algorithm}[h]
\caption{\textsc{BuildValidator}}
\KwInput{Problem spec $\mathcal{S}$}
\KwOutput{Selected Validator $V^\star$}
$\mathcal{E} \leftarrow$ \textsc{LLM.GenerateEvalCases}$(
  \mathcal{S};~N{=}40,~10~\text{valid},~30~\text{near-valid})$

$\{V_1,V_2,V_3\} \leftarrow$ \textsc{LLM.EmitValidators}$(\mathcal{S},~K{=}3)$\;
\For{$V$ in $\{V_1,V_2,V_3\}$}{
  $\text{score}(V) \leftarrow \sum_{(x,\text{label}) \in \mathcal{E}} [V(x){=}\text{label}]$\;
}
$V^\star \leftarrow \arg\max_{V \in \{V_1,V_2,V_3\}} \text{score}(V)$\;
\Return $V^\star$\;
\end{algorithm}

The foundation of our framework is a highly robust Validator, responsible for rejecting any input that violates the problem's explicit constraints. To construct it, we first guide the LLM to generate a targeted suite of 40 test cases. This suite is strategically composed of 10 valid inputs and 30 near-valid illegal inputs, i.e., cases that become valid after minor modifications. We present an example in Appendix~\ref{appendix:test_case_gen_case_study}. These carefully crafted and validated edge cases are designed to rigorously probe the Validator's robustness against subtle constraint violations. With this evaluation set established, we then prompt the LLM to produce three distinct candidate Validator programs. Each candidate is subsequently benchmarked against the 40 cases. The program that correctly classifies the highest number of these valid and invalid test cases is selected as the definitive Validator for all subsequent stages of the framework, ensuring a strong guard against malformed data.

\subsection{Generator}

\begin{algorithm}[h]
\caption{\textsc{BuildGeneratorSuite}}
\KwInput{$\mathcal{S}$, Validator $V^\star$}
\KwOutput{Final test set $\mathcal{T}$}
$\mathcal{G}_1 \leftarrow$ \textsc{ExhaustiveSmall}$(\mathcal{S})$ \Comment*[r]{small-scale enumeration coverage}
$\mathcal{G}_2 \leftarrow$ \textsc{RandomExtreme}$(\mathcal{S})$ \Comment*[r]{random + extreme: overflows, precision, hash collisions, etc.}
$\mathcal{G}_3 \leftarrow$ \textsc{TLEInducing}$(\mathcal{S})$ \Comment*[r]{worst-case structures inducing TLE}
$\mathcal{U} \leftarrow \mathcal{G}_1 \cup \mathcal{G}_2 \cup \mathcal{G}_3$\;
$\mathcal{U} \leftarrow \{x \in \mathcal{U} ~|~ V^\star(x){=}\textsf{valid}\}$ \Comment*[r]{Validator filters invalid inputs}
$\mathcal{U} \leftarrow$ \textsc{DedupBySignature}$(\mathcal{U})$ \Comment*[r]{hash/normalization-based deduplication}
$\mathcal{U} \leftarrow$ \textsc{BalanceBuckets}$(\mathcal{U};~\text{size/structure/hardness})$\;
$\mathcal{T} \leftarrow$ \textsc{SampleWithCoverage}$(\mathcal{U};~\text{target size})$\;
\Return $\mathcal{T}$\;
\end{algorithm}

After ensuring input validity through a precise Validator, the Generator's core task becomes maximizing test coverage. By creating challenging test cases, it can more effectively identify incorrect or inefficient solutions, thereby reducing FPR. To enhance the coverage of our test cases, we adopt strategies across three distinct dimensions.

\begin{itemize}[leftmargin=9pt]
\item \textbf{Small Data Exhaustion.} For problems with small constraints or those featuring multiple test points, as are common on platforms like Codeforces and AtCoder, we generate inputs that exhaustively explore all permutations and combinations of small-scale data. This strategy ensures complete coverage of fundamental boundary conditions and corner cases.

\item \textbf{Randomized and Extreme Data.} We generate large-scale random inputs to stress-test solutions. This includes pushing integer types to their limits to trigger overflows, testing floating-point precision, and probing for out-of-bounds array access. Drawing from contest experience, we also adversarially design hack cases, such as those intended to make common greedy algorithms fail or to induce hash collisions.

\item \textbf{TLE-Inducing Data.} To specifically address the common issue of solutions with incorrect time complexity passing on weak test cases, we construct inputs with specific structures designed to maximize the computational load for certain algorithms. This ensures that only solutions meeting the intended time complexity requirements can pass, effectively catching false positive verdicts caused by insufficient timing pressure.
\end{itemize}

\subsection{Checker}

\begin{algorithm}[h]
\caption{\textsc{BuildChecker}}
\KwInput{$\mathcal{S}$, reference solution $\mathcal{R}$, Validator $V^\star$}
\KwOutput{Selected Checker $C^\star$}
$\mathcal{Q} \leftarrow$ \textsc{LLM.GenerateCheckerScenarios}$(
  \mathcal{S}, \mathcal{R};~N{=}40)$\;
\ForEach{$q \in \mathcal{Q}$}{
  \If{$V^\star(q.\text{input}) \neq \textsf{valid}$}{ remove $q$ } \Comment*[r]{ensure input legality}
}
$\{C_1,C_2,C_3\} \leftarrow$ \textsc{LLM.EmitCheckers}$(\mathcal{S},~K{=}3)$\;
\For{$C$ in $\{C_1,C_2,C_3\}$}{
  $acc(C) \leftarrow \frac{1}{|\mathcal{Q}|}\sum_{q\in\mathcal{Q}}
  [\,C(q.\text{input},~q.\text{contestant\_out},~q.\text{ref\_out}){=}q.\text{verdict}\,]$\;
}
$C^\star \leftarrow \arg\max_C acc(C)$\;
\Return $C^\star$\;
\end{algorithm}

The Checker is responsible for determining the final verdict of a submission by comparing the output with the answer output provided by the reference solution. To ensure that the Checker achieves a similar accuracy as the Validator, we adopt a construction method analogous to that used for the Validator. First, we prompt the LLM to generate 40 distinct test cases. Each test case consists of a valid input, an output that appears reasonable but might contain mistakes, the correct output provided by the reference solution, and the expected verdict (e.g., Accepted or Wrong Answer). The validity of all inputs is guaranteed by the previously established Validator. Next, the LLM generates three candidate Checker programs. We evaluate each candidate Checker against these 40 test cases to assess their ability to produce accurate verdicts. Finally, the Checker program that performs best is selected to be part of the final test data package.

\subsection{Interactor}
\vspace{-6pt}
\begin{algorithm}[h]
\caption{\textsc{BuildInteractor}}
\KwInput{$\mathcal{S}$, reference solution $\mathcal{R}$, Validator $V^\star$}
\KwOutput{Selected Interactor $I^\star$}
$\mathcal{M} \leftarrow$ \textsc{LLM.Mutate}$(
  \mathcal{R};~\text{small logical edits: }{<}{>}\text{/}\leq\text{/}\geq~\text{swap, off-by-one, missing checks, wrong tie-breaks, RNG misuse, etc.})$\;
$\{I_1,I_2,I_3\} \leftarrow$ \textsc{LLM.EmitInteractors}$(\mathcal{S},~K{=}3)$\;
\For{$I$ in $\{I_1,I_2,I_3\}$}{
  $p \leftarrow [\,\textsc{Simulate}(I,\mathcal{R},V^\star){=}\textsf{Accepted}\,]$\;
  $f \leftarrow \sum_{m \in \mathcal{M}} [\,\textsc{Simulate}(I,m,V^\star){=}\textsf{Rejected}\,]$\;
  $\text{score}(I) \leftarrow (p,~f)$ \Comment*[r]{lexicographic: prioritize passing the true solution, then maximize discrimination}
}
$I^\star \leftarrow \arg\max_I \text{score}(I)$\;
\Return $I^\star$\;
\end{algorithm}

Our framework introduces a novel, fully automated approach for generating test data for interactive problems, a previously unaddressed challenge (e.g., in CodeContests+~\cite{wang2025codecontests+}). The core innovation lies in a mutant-based discrimination process. We begin by prompting the LLM to perform several small but critical logical modifications to the provided reference solution, thereby creating a set of slightly incorrect mutant programs. Examples are shown in Appendix~\ref{appendix:wrong_solution}. These mutants serve as challenging foils for the Interactor. The LLM then generates three candidate Interactor programs. The crucial selection criterion is identifying the Interactor that most effectively distinguishes the correct, unmodified reference solution from the flawed mutant versions during a simulated interaction shown in Appendix~\ref{appendix:interactor}. The Interactor that successfully passes the reference solution while failing the maximum number of mutants is chosen, proving its ability to robustly probe for the specific logic required by the problem.

\section{Benchmarking Test Case Robustness}
\label{sec:bench}

\begin{table}[t]
\centering
\caption{Performance comparison on the 7538-problem benchmark. The evaluation is performed on a set of 195,988 human submissions randomly taken from the CodeContests dataset, where each problem has 26 submissions, 50\% of which are correct and 50\% are incorrect. Results are reported with 95\% confidence intervals. The results for AutoCode are obtained using \texttt{o3}.}
\label{tab:main_results}
\resizebox{0.88\textwidth}{!}{
\begin{tabular}{lccc}
\toprule
\textbf{Method} & \textbf{Consistency (\%)~($\uparrow$)} & \textbf{FPR (\%)~($\downarrow$)} & \textbf{FNR (\%)~($\downarrow$)} \\
\midrule
CodeContests~\cite{li2022competition}   & 72.9 $\pm$ 0.2 & 7.7 $\pm$ 0.2  & 46.3 $\pm$ 0.3 \\
CodeContests+~\cite{wang2025codecontests+} & 79.9 $\pm$ 0.2 & 8.6 $\pm$ 0.2  & 31.6 $\pm$ 0.3 \\
TACO~\cite{li2023taco}                  & 80.7 $\pm$ 0.2 & 11.5 $\pm$ 0.2 & 26.9 $\pm$ 0.3 \\
HardTests~\cite{he2025hardtests}        & 81.0 $\pm$ 0.2 & 12.1 $\pm$ 0.2 & 25.8 $\pm$ 0.3 \\
\midrule
\model~(Ours)                           & \textbf{91.1 $\pm$ 0.1} & \textbf{3.7 $\pm$ 0.1} & \textbf{14.1 $\pm$ 0.2} \\
\bottomrule
\end{tabular}
}
\end{table}

\begin{table}[t]
\label{tab:ablation}
\centering
\caption{Ablation study results on the 720-problem benchmark. For each problem, the evaluation uses 33 submissions generated by different LLMs, 25\% of which are accepted. This setup is particularly relevant for reinforcement learning applications. Each row shows performance after removing a single component from the full framework. All results in this table are generated by \texttt{GPT-5-High}.}
\label{tab:ablation}
\resizebox{0.98\textwidth}{!}{
\begin{tabular}{lccc}
\toprule
\textbf{Configuration} & \textbf{Consistency (\%)~($\uparrow$)} & \textbf{FPR (\%)~($\downarrow$)} & \textbf{FNR (\%)~($\downarrow$)} \\
\midrule
w/o Generator Strategy 1 (Exhaustive) & 98.4 $\pm$ 0.2 & 1.7 $\pm$ 0.2 & 1.3 $\pm$ 0.3 \\
w/o Generator Strategy 2 (Random/Extreme) & 98.4 $\pm$ 0.2 & 1.6 $\pm$ 0.2 & 1.3 $\pm$ 0.3 \\
w/o Generator Strategy 3 (TLE) & 98.6 $\pm$ 0.2 & 1.4 $\pm$ 0.2 & 1.3 $\pm$ 0.3 \\
\midrule
w/o Prompt Optimization & 98.0 $\pm$ 0.2 & 1.8 $\pm$ 0.2 & 2.9 $\pm$ 0.4 \\
\midrule
\text{\model~Full Framework} & \textbf{98.7 $\pm$ 0.1} & \textbf{1.3 $\pm$ 0.2} & \textbf{1.2 $\pm$ 0.3} \\
\bottomrule
\end{tabular}
}
\end{table}

To rigorously evaluate our test case generation framework, we establish two distinct benchmarks. The primary benchmark consists of 7538 problems derived from the intersection of well-known existing datasets: CodeContests+~\cite{wang2025codecontests+}, CodeContests, HardTests~\cite{he2025hardtests}, and TACO~\cite{li2023taco}. Notably, this large-scale set does not contain interactive problems, and due to the filtering inherent in these datasets, its average difficulty for test data generation is slightly lower than a typical Codeforces round.

To address this and test our system under more challenging, real-world conditions, we create a second benchmark of 720 recent, rated problems from Codeforces. This set is completely unfiltered, including notoriously difficult-to-handle interactive problems and those requiring complex, structured test data. We are unable to evaluate prior methods on this newer benchmark as their data generation codebases are not publicly available.

Our evaluation is based on three key metrics. Consistency measures the overall percentage of agreement between the verdicts from our tests and the official judgments. We further dissect disagreements into two critical error rates. The FPR is defined as the proportion of officially incorrect solutions that are erroneously accepted by our generated tests. Conversely, the FNR is the proportion of officially correct solutions that are erroneously rejected by our tests.

\begin{figure}[t]
    \centering
    \includegraphics[width=0.99\linewidth]{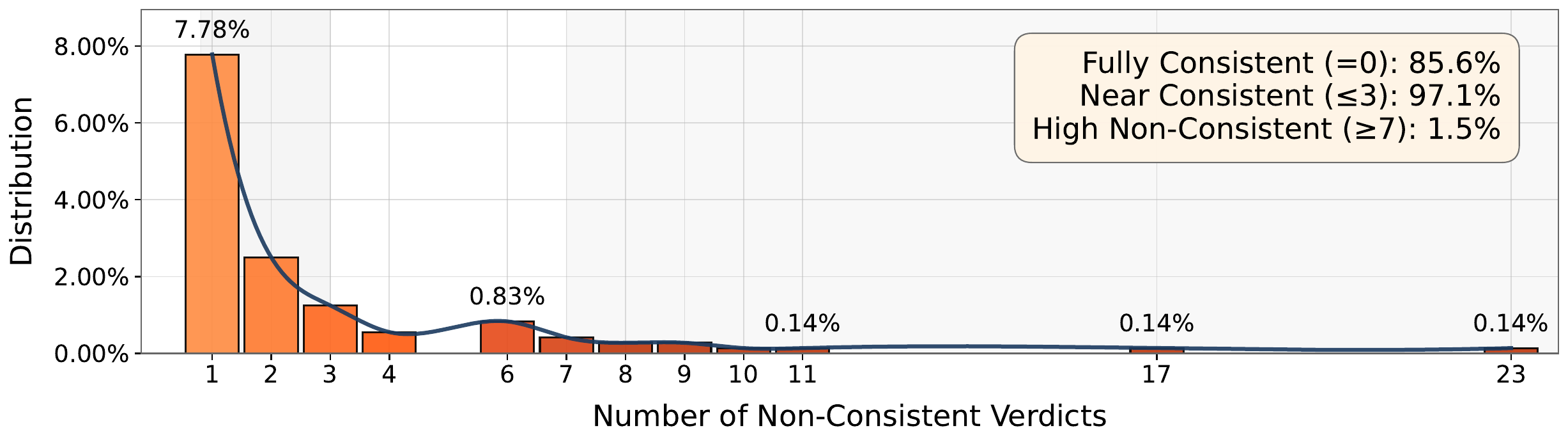}
    \caption{\textbf{Distribution of Non-Consistent Verdicts.} Each problem in the 720-problem benchmark is evaluated using 33 submissions generated by different LLMs. The verdicts for 85.6\% of the problems are consistent with the official judgments, and 97.1\% have three or fewer inconsistencies.}
    \label{fig:pdf}
\end{figure}

\vspace{-0.5em}
\paragraph{Comparison with other baselines.}
We evaluate~\model~on the benchmark of 7538 problems against four leading baselines. As detailed in Table~\ref{tab:main_results}, our framework achieves 91.1\% consistency with official judgments. This marks a significant leap over existing methods which don't surpass 81.0\%. Critically, \model~substantially reduces the FPR to just 3.7\% and the FNR to 14.1\%, representing $\approx 50\%$ decrease in both metrics over the current state-of-the-art. Figure~\ref{fig:pdf} shows the error verdicts distribution, showing that most of the problems are consistent with ground truth verdicts.

To further test our system's robustness, we curate a more challenging benchmark of 720 recent, unfiltered Codeforces problems, including complex interactive tasks. As shown in Table~\ref{tab:ablation}, \model~maintains its exceptional performance, achieving 98.7\% consistency. This result validates our method's effectiveness on modern, difficult problems where prior methods could not be evaluated.

\vspace{-0.5em}
\paragraph{Ablation studies.}

We further conduct an ablation study to determine how each part of AutoCode affects overall performance. The complete AutoCode framework sets a strong baseline, achieving 98.7\% consistency with official judgments, along with a 1.3\% FPR and a 1.2\% FNR. Prompt optimization turns out to be especially important; removing it drops consistency to 98.0\% and more than doubles the FNR to 2.9\%. The three different generator strategies also significantly complement each other. Removing either the exhaustive or random/extreme strategy raises the FPR to 1.7\% and 1.6\%, respectively, highlighting their role in catching flawed solutions.

Validator and Checker selection tests are also crucial. These tests pick the most reliable output from several model-generated candidates, helping weaker models avoid logical errors. Removing these tests increases the FNR to 1.3\% and 1.4\%. A good Validator prevents correct solutions from being rejected due to bad inputs, while a strong Checker accurately evaluates outputs with complex judgment rules. Even small improvements (0.1\%) matter, as they help address problems that are very difficult to generate test cases for.

\vspace{-0.5em}
\section{Problem Generation}
\label{sec:problem_generation}
\vspace{-0.5em}

Our novel problem generation framework builds upon the robust test generation framework described in Section~\ref{sec:test_case_gen} and shown in Figure~\ref{fig:teasor}, but introduces a critical dual-verification protocol to ensure correctness without human intervention. 

Each generated problem is graded on a 6-level scale, judged by top human competitive programmers. We interviewed 8 human expert problem setters, all of whom report that they often build on a specific existing problem when authoring new problems. By adding, removing, or modifying certain conditions of such a ``seed problem,'' they create new and often more difficult problems that require novel insights. Inspired by their insights, our approach begins by selecting a random Codeforces problem (with difficulty rating less than 2200) as a ``seed problem.'' The LLM is tasked with generating a new problem by adding, deleting, or modifying certain conditions from this seed problem, along with an efficient reference solution (std.cpp) and a brute-force solution (brute.cpp). brute.cpp usually has higher time complexity but is very unlikely to be incorrect, so we leverage it to stress-test the validity of the problem. Using our enhanced test case generation technique, we construct a comprehensive set of test data that fully covers small cases. Both brute.cpp and std.cpp are then executed on this dataset. A problem is only deemed correct if, for every test case, both programs’ outputs (where the brute-force solution may legitimately fail to finish due to timeout) are pairwise validated by the checker as consistent answers and outputs. This dual-verification protocol, where brute.cpp serves as the initial ground truth, and the validated reference solution undergoes an additional full test generation cycle, successfully filters out 27\% of error-prone problems, raising the correctness rate of LLM-provided reference solutions from 86\% to 94\%. 

\begin{figure}[t]
    \centering
    \includegraphics[width=0.98\linewidth]{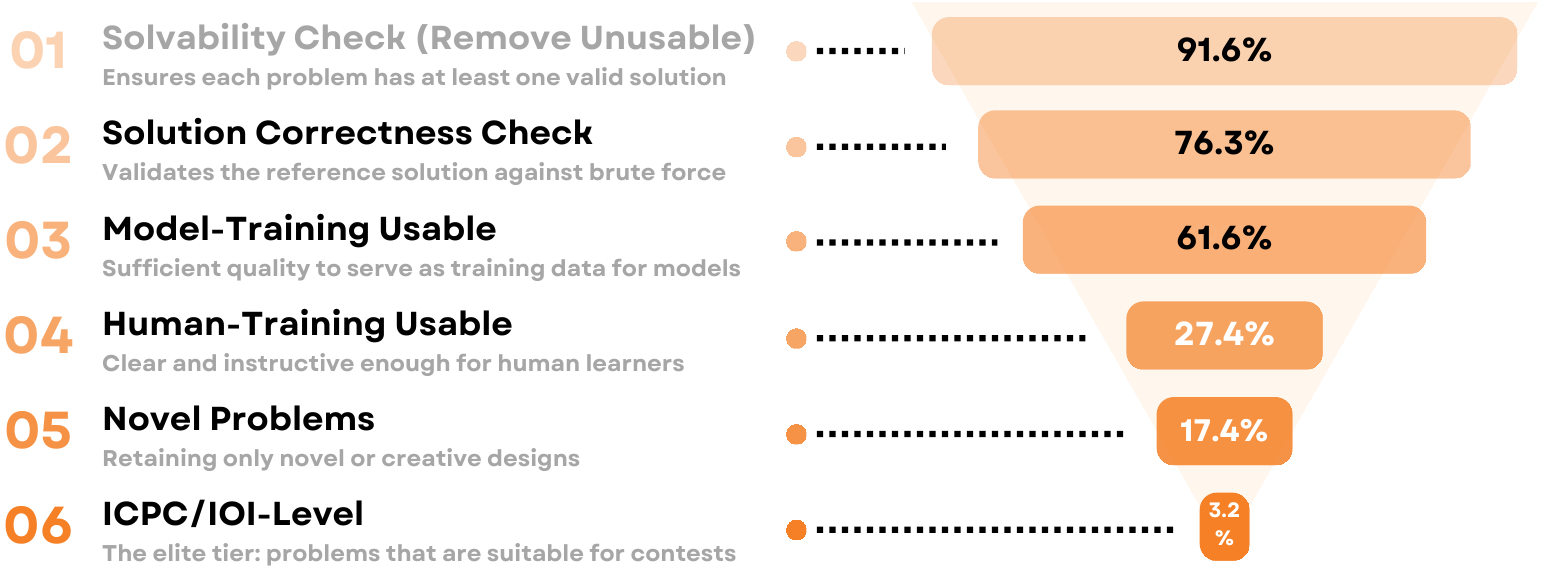}
    \caption{\textbf{\model}~is capable of automatically generating novel competitive programming problems. After careful examination and filtering by multiple human experts, 61.6\% achieve a quality level suitable for large language model training, and 3.2\% are considered good enough to serve as problems for ICPC/IOI-level contests. Level 1 and 2 correspond to automatic correctness checks performed by the system, while level 3 through 6 involve human evaluation to differentiate among finer levels. It is noted that only the Level 1 questions are wrong, while the Level 2 -- 6 questions are all correct and usable, with the only difference being their quality. Detailed grading rubrics are in Appendix~\ref{appendix:grade}.}

    \label{fig:problem_level}
    \vspace{-12pt}
\end{figure}

After filtering, over 80\% of the problems are annotated as having sufficient quality to serve as training data for models, and 23\% of the problems have novel or creative designs involved. 
We present the detailed rubrics and the score distribution in Figure~\ref{fig:problem_level}. In the following, we summarize several key findings regarding the performance of LLMs in problem generation. We also present the human expert grading criteria in Appendix~\ref{appendix:grade} and one example under ICPC/IOI-level in Appendix~\ref{appendix:generated_problem}. 

\finding{1}{LLMs can generate solvable problems that they themselves are unable to solve.}

LLMs can generate \textit{solvable} problems that they themselves are \textit{unable to solve}. In our experiments, about 4.2\% of the problems fall into this category. In other words, the model is indeed capable of creating logically sound problems that can be solved by other models or by humans, but due to limitations in its own reasoning ability, it fails to solve them correctly. 
Ideally, these problems could serve as sources for model self-improvement, and we think this is a very interesting phenomenon that is worth further investigation. We provide an example in the Appendix~\ref{appendix:fail}.

\finding{2}{LLMs tend to create new problems by combining existing problem frameworks and emphasizing knowledge and implementations.}

LLMs tend to concatenate, combine, or embed existing algorithmic knowledge or tricks into established problem frameworks, rather than proposing entirely new problem models that require original solution strategies. This form of recombinational innovation primarily reflects reuse of existing knowledge rather than expansion of creative thinking. Such a tendency highlights a fundamental divergence between humans and LLMs in defining novelty: while humans emphasize originality in modes of reasoning and problem-solving ideas, LLMs rely more heavily on recombination and reconfiguration of pre-existing knowledge. Also, the generated problems often place greater emphasis on assessing specific algorithmic knowledge or demanding complex implementation details, while less often rely on clever design and subtle observation.

\finding{3}{Novel problems tend to have larger difficulty gain over seed problems, and the generated problems have the highest quality when the corresponding seed problem is moderately difficult.}

We observe that the difficulty shifts induced by LLMs during problem adaptation follow a systematic pattern rather than occurring randomly. On average, adapted problems become approximately 334 Elo score harder; those judged as ``novel'' show an average increase of 498 score, whereas non-novel problems increase by only about 108 score. High-quality problems are predominantly generated when the original seed problem lies in the moderately high difficulty range. For overly difficult seed problems, the LLM has limited room to introduce effective modifications, resulting in minimal difficulty gains and insufficient novelty. Conversely, for overly easy seed problems, even after an average increase of 334 score, the absolute difficulty remains too low to meet high-quality standards. Notably, around 5\% of generated problems fall into a critical zone, with pass@1 scores between 0.1 and 0.5, meaning the model sometimes succeeds and sometimes fails. These boundary cases present a valuable opportunity for constructing high-quality self-play datasets, enabling models to enhance their capabilities through repeated attempts at solving such borderline problems.

\finding{4}{Human experts and LLMs show almost no correlation in their judgment of problem quality and novelty.}

\begin{figure}[t]
    \centering
    \includegraphics[width=0.98\linewidth]{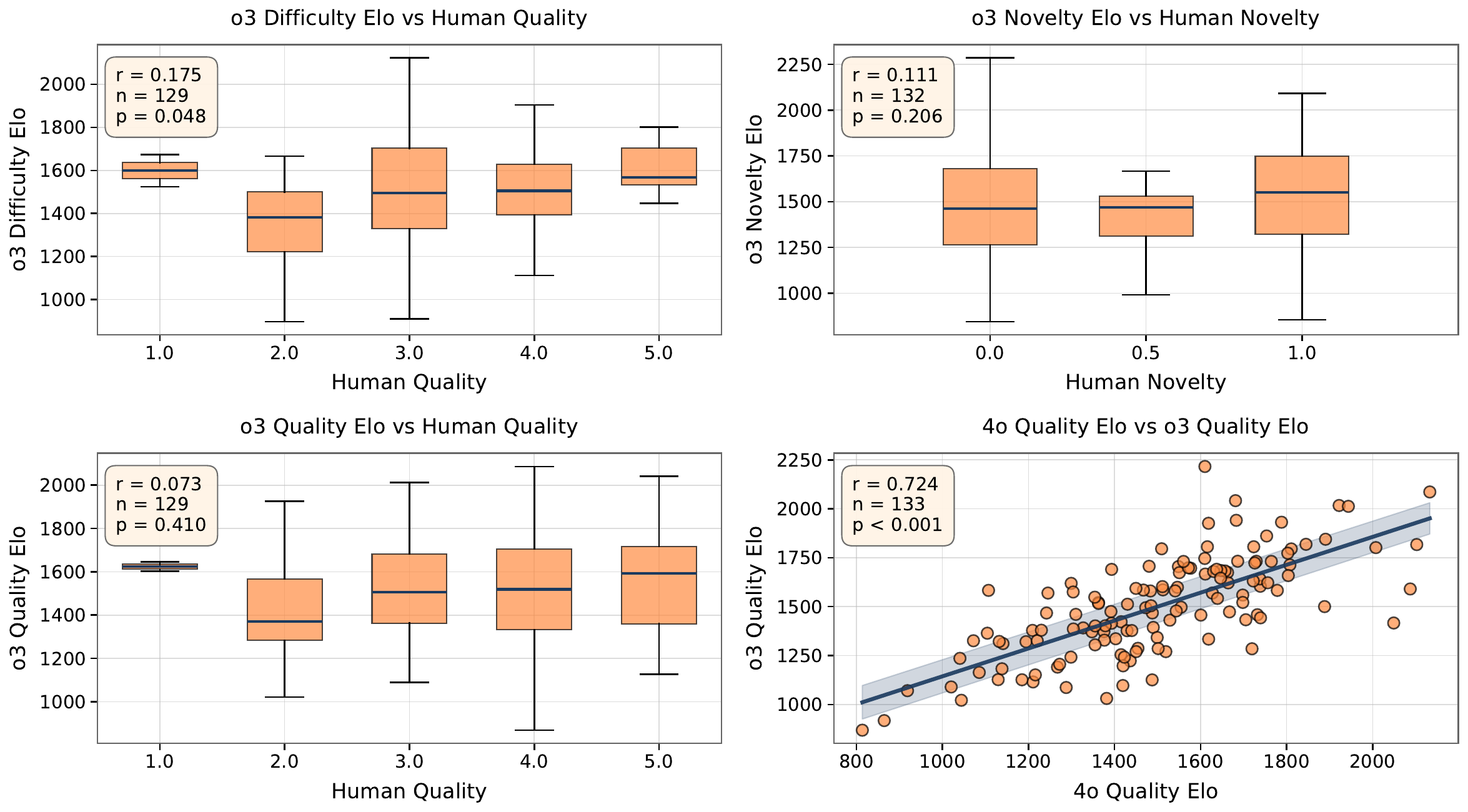}
    \caption{Correlations between human experts and LLM judgments on generated problems. The charts illustrate a significant gap between human and LLM perceptions of quality and novelty (Finding 4), despite high inter-LLM agreement (bottom right). LLM-predicted difficulty shows a weak positive correlation with human-rated quality (top left), suggesting it can serve as a noisy proxy for quality estimation (Finding 5).}
    \label{fig:correlation}
    \vspace{-12pt}
\end{figure}

We identify a significant divergence between humans and LLMs in their judgment of problem quality and novelty. To quantify this discrepancy, we employ an Elo Rating scheme to assess LLM judgments, following the methodology described in ~\cite{zhou2024evaluating}. The correlation coefficients between the \texttt{o3} and human experts are only 0.07 for quality and 0.11 for novelty, indicating a substantial misalignment between the model’s internal evaluation standards and human expert criteria. Interestingly, both humans and LLMs demonstrate high within-group consistency in terms of quality: correlations among human experts reach 0.71, while \texttt{GPT-4o} and \texttt{o3} show a correlation of 0.72. These findings suggest that relying solely on LLMs to self-evaluate the quality of their generated problems is inadequate, and more sophisticated evaluation mechanisms are required to align with human preferences as also shown in Figure~\ref{fig:correlation}.

\finding{5}{Difficulty of the generated problem and difficulty gain over the seed problem serve as a better indicator of problem quality than LLM self-evaluations.}

While LLMs are unreliable in directly assessing problem quality, leveraging their predictions of problem difficulty provides a more effective indirect measure. First, difficulty is the strongest predictor of quality: correlations between human-assigned quality scores and both absolute problem difficulty and difficulty gain reach as high as 0.60. Second, difficulty gain outperforms novelty as an indicator. As a continuous variable, it captures richer information, since problems that exhibit sufficiently large difficulty increases tend to incorporate new ideas or more complex combinations, thereby manifesting novelty. Finally, LLMs possess a certain ability to estimate problem difficulty; by exploiting this capacity, difficulty can be used as a proxy signal to indirectly predict problem quality, achieving correlations of around 0.18 as shown in Figure~\ref{fig:correlation}.

\vspace{-6pt}
\section{Conclusion}

In this work, we introduce~\model, a closed-loop multi-role framework that leverages LLMs as problem setters for competitive programming. By coupling a Validator-Generator-Checker (and Interactor) framework with a dual-verification protocol, \model~achieves state-of-the-art reliability in test case generation and extends beyond prior methods to generate entirely new, contest-quality problems. Extensive experiments on over 7,500 problems and recent Codeforces benchmarks demonstrate that~\model~substantially reduces both false positives and false negatives, achieving more than 98\% agreement with official judgments and successfully producing novel problems validated by expert programmers. Beyond test generation, our analysis sheds light on both the strengths and weaknesses of LLMs in creative problem authoring. While models excel at recombination of algorithmic knowledge, they struggle to introduce truly novel reasoning paradigms or flawless sample design. Nevertheless, we show that difficulty and difficulty gain serve as reliable proxy signals for problem quality, providing a scalable path toward self-play.

\clearpage
\bibliography{references}

\clearpage
\beginsupplement
\begin{center}
     \Large\textbf{Appendix}
\end{center}

\noindent The appendix is structured as follows:
\begin{itemize}[leftmargin=9pt]
\setlength{\itemsep}{2pt}
\item Preliminaries and definitions of key concepts and terms in Section~\ref{preliminaries}.
\item Example of Competition-Grade level generated problem in Section~\ref{appendix:generated_problem}.
\item Example of generated problem that LLM itself can't solve in Section~\ref{appendix:fail}.
\item Example of valid inputs and near-valid illegal inputs in Section~\ref{appendix:test_case_gen_case_study}.
\item Example of incorrect mutant programs in Section~\ref{appendix:wrong_solution}.
\item Human experts grading policy on LLM generated problems in Section~\ref{appendix:grade}.
\item Pseudo code of simulator in interactor in Section~\ref{appendix:interactor}.
\item Limitations and future work in Section~\ref{appendix:limitation}.
\end{itemize}

\section{Preliminaries}
\label{preliminaries}
In this section, we provide background information on the judgment process in competitive programming, define key terminologies used throughout the paper, and detail the core components of the \model framework.

\subsection{Competitive Programming Solution Judgment}

In competitive programming, solutions are judged by executing them against a series of test cases. Each test case consists of an input and a ground-truth output. The system compares the program's output for each input against the ground-truth answer. A solution is only considered correct if it passes all test cases while executing within the problem's time and memory constraints. The outcome of the judgment process is a single verdict. The most common verdicts are:

\begin{itemize}
\item {\bf Accepted (AC):} The solution passed all test cases within the given time and memory limits. This is the only successful outcome.

\item {\bf Wrong Answer (WA):} The solution produces outputs that fail on at least one test case.

\item {\bf Time Limit Exceeded (TLE):} The solution exceeds the time limit on at least one test case.

\item {\bf Memory Limit Exceeded (MLE):} The solution exceeds the memory limit on at least one test case. 

\item {\bf Runtime Error (RE):} The solution terminates abnormally due to an error like a segmentation fault or division by zero. 
\end{itemize}

\subsection{Key Terminologies}
\paragraph{Codeforces.} A popular online platform that hosts competitive programming contests. The difficulty of problems on this platform is often used as a benchmark.

\paragraph{Elo rating for problem difficulty.} On Codeforces, each problem has a numerical rating assigned to a problem to quantify its difficulty. It's calibrated based on the performance of contestants on this problem during the competition. In general, a difficulty rating of $\leq 2000$ is deemed as easy to medium range, $(2000, 3000]$ is deemed as medium to hard range, and $\geq 3000$ is deemed as very hard~\cite{zheng2025livecodebench}.

\paragraph{Generator.} A program that produces the input files for the test cases. A robust set of generators is needed to create a diverse set of test cases that cover edge cases, average cases, and worst-case scenarios designed to challenge the time or memory complexity of incorrect algorithms.

\paragraph{Validator.} A program that verifies whether a test case conforms to the constraints described in the problem statement (e.g., $1 \leq N \leq 1000$, the given graph is a tree). 

\paragraph{Checker.} A program that compares a contestant's output to the correct answer and issuing a verdict. Many problems require custom Checkers beyond simple text comparisons, including
\begin{itemize}
\item {\bf Problems with multiple correct solutions}: For example, in a constructive problem asking for any valid path in a graph, the Checker must verify the validity of the contestant's proposed path, not just match it to one specific example.

\item {\bf Problems with floating-point outputs}: When the answers are floating-point numbers, answers are accepted if they are within a certain absolute or relative error tolerance (e.g., $10^{-6}$).

\item {\bf Problems where output format is flexible}: For instance, some problems that require contestants to output ``Yes'' or ``No'' may also accept ``YES'' or ``NO'' as correct answers.
\end{itemize}

\paragraph{Interactor.} A specialized component is used for interactive problems. In an interactive problem, a contestant's solution does not receive all input at once. Instead, it engages in a real-time dialogue with the judge's program (the Interactor). The solution makes a series of queries to the Interactor, receives responses, and use the information gathered from this dialogue to determine the final answer. A classic example is a guessing game where the Interactor knows a secret number, and the solution must find it by making queries and receiving ``higher'' or ``lower'' as responses.
\section{Example of LLM Generated Problem: Row--Column Portal}
\label{appendix:generated_problem}

\textbf{Time limit per test:} 2 seconds \\
\textbf{Memory limit per test:} 256 megabytes

You are given a binary grid $A$ with $n$ rows and $m$ columns ($0 =$ empty, $1 =$ obsidian).

Fix two integers $a$ and $b$ with $5 \leq a \leq n$ and $4 \leq b \leq m$. A sub-rectangle $M$ of size $a \times b$ (with rows $r..r+a-1$ and columns $c..c+b-1$) is called a \textbf{portal} if and only if:
\begin{itemize}
    \item Every cell on the border of $M$ is $1$ \textbf{except} the four corner cells (which can be either $0$ or $1$).
    \item Every strictly interior cell of $M$ is $0$.
\end{itemize}

You are allowed to choose \textbf{one} sub-rectangle of size $a \times b$ and then perform the following operation any number of times \textbf{inside that chosen sub-rectangle only}:
\begin{itemize}
    \item Pick a row of the sub-rectangle and flip all its cells ($0 \leftrightarrow 1$), or
    \item Pick a column of the sub-rectangle and flip all its cells ($0 \leftrightarrow 1$).
\end{itemize}

Flips affect only the chosen $a \times b$ sub-rectangle (not the rest of the grid).  
Your task is to find the \textbf{minimum number of flips} (rows + columns) needed to turn \textbf{some} $a \times b$ sub-rectangle into a portal. If it is impossible for every placement, print $-1$.

\subsection*{Input}
The first line contains an integer $t$ ($1 \leq t \leq 2\cdot 10^5$) — the number of test cases.

For each test case:
\begin{itemize}
    \item The first line contains four integers $n, m, a, b$ ($5 \leq a \leq n \leq 2000$, $4 \leq b \leq m \leq 2000$).
    \item Then follow $n$ lines, each a string of length $m$ consisting of characters 0 and 1.
\end{itemize}

It is guaranteed that the sum of $n \cdot m$ over all test cases does not exceed $4 \cdot 10^5$.

\subsection*{Output}
For each test case, output a single integer — the minimal number of flips needed, or $-1$ if impossible.

\subsection*{Examples} 

\textbf{Input}
\begin{verbatim}
1
5 4 5 4
1111
1001
1001
1001
1111
\end{verbatim}

\textbf{Output}
\begin{verbatim}
0
\end{verbatim}

\textbf{Input}
\begin{verbatim}
1
6 6 5 4
010101
111000
001010
111111
000000
101010
\end{verbatim}

\textbf{Output}
\begin{verbatim}
5
\end{verbatim}

\subsection*{Accepted Solution (C++)}

\begin{lstlisting}[style=cpp]
#include <bits/stdc++.h>
using namespace std;

// 2D prefix sum over a matrix of ints (0/1).
struct Pref2D {
    int n, m;
    vector<vector<int>> ps; // (n+1) x (m+1)
    Pref2D() {}
    Pref2D(int n_, int m_) : n(n_), m(m_), ps(n_+1, vector<int>(m_+1, 0)) {}
    void build(const vector<vector<int>>& a) {
        for (int i = 0; i < n; ++i) {
            int rowSum = 0;
            for (int j = 0; j < m; ++j) {
                rowSum += a[i][j];
                ps[i+1][j+1] = ps[i][j+1] + rowSum;
            }
        }
    }
    int sum(int r1, int c1, int r2, int c2) const {
        if (r1 > r2 || c1 > c2) return 0;
        return ps[r2+1][c2+1] - ps[r1][c2+1] - ps[r2+1][c1] + ps[r1][c1];
    }
};

int main() {
    ios::sync_with_stdio(false);
    cin.tie(nullptr);
    int t;
    if (!(cin >> t)) return 0;
    while (t--) {
        int n, m, a, b;
        cin >> n >> m >> a >> b;
        vector<string> s(n);
        for (int i = 0; i < n; ++i) cin >> s[i];

        vector<vector<int>> A(n, vector<int>(m, 0));
        for (int i = 0; i < n; ++i)
            for (int j = 0; j < m; ++j)
                A[i][j] = (s[i][j] == '1');

        vector<vector<int>> rowPS(n, vector<int>(m+1, 0));
        vector<vector<int>> colPS(n+1, vector<int>(m, 0));
        for (int i = 0; i < n; ++i) {
            for (int j = 0; j < m; ++j) {
                rowPS[i][j+1] = rowPS[i][j] + A[i][j];
                colPS[i+1][j] = colPS[i][j] + A[i][j];
            }
        }

        int NQn = max(0, n-1), NQm = max(0, m-1);
        vector<vector<int>> Q(NQn, vector<int>(NQm, 0));
        for (int i = 0; i+1 < n; ++i) {
            for (int j = 0; j+1 < m; ++j) {
                Q[i][j] = A[i][j] ^ A[i+1][j] ^ A[i][j+1] ^ A[i+1][j+1];
            }
        }
        Pref2D Qps(NQn, NQm);
        Qps.build(Q);

        long long best = (long long)1e18;
        int h = a, w = b;

        for (int r = 0; r + h <= n; ++r) {
            for (int c = 0; c + w <= m; ++c) {
                int rq1 = r, cq1 = c, rq2 = r + h - 2, cq2 = c + w - 2;
                int total = Qps.sum(rq1, cq1, rq2, cq2);
                if (h >= 2 && w >= 2) {
                    total -= Q[rq1][cq1];
                    total -= Q[rq1][cq2];
                    total -= Q[rq2][cq1];
                    total -= Q[rq2][cq2];
                }
                if (total != 0) continue;

                int pivotRow = r + 1, pivotCol = c + 1;
                int p = A[pivotRow][pivotCol];

                int innerColsOnes = rowPS[pivotRow][c + w - 1] - rowPS[pivotRow][c + 1];
                int leftEnd = A[pivotRow][c] ^ 1;
                int rightEnd = A[pivotRow][c + w - 1] ^ 1;
                int S_cols = innerColsOnes + leftEnd + rightEnd;

                int L = r + 1, U = r + h - 2;
                int len = max(0, U - L + 1);
                int onesInSeg = (len ? (colPS[U+1][pivotCol] - colPS[L][pivotCol]) : 0);
                int unequalInterior = (p == 0 ? onesInSeg : (len - onesInSeg));

                int d_top = A[r][pivotCol] ^ p;
                int d_bot = A[r + h - 1][pivotCol] ^ p;
                int S_rows = unequalInterior + 2 - d_top - d_bot;

                int S0 = S_rows + S_cols;
                int flips = min(S0, h + w - S0);
                best = min(best, (long long)flips);
            }
        }

        if (best == (long long)1e18) cout << -1 << "\n";
        else cout << best << "\n";
    }
    return 0;
}
\end{lstlisting}

\section{Example of LLM Generated Problem that Itself Cannot Solve: IT Restaurants — Distance-2 Rivalry}
\label{appendix:fail}

\textbf{Time limit per test:} 2 seconds \\
\textbf{Memory limit per test:} 256 megabytes

City $N$ still has a tree road network with $n$ junctions ($n-1$ roads, connected, undirected). The mayor again wants to place restaurants of two rival IT networks: iMac Donalds and Burger Bing. Each junction can host at most one restaurant.

This time, the owners impose a stricter rule set:

\begin{itemize}
    \item Rivalry (distance-1): Two neighboring junctions (joined by a road) cannot host restaurants of different networks.
    \item Brand dilution (distance-2): Two junctions at distance exactly 2 (i.e., they have a common neighbor) cannot host restaurants of the same network.
\end{itemize}

Both networks must build at least one restaurant. The mayor wants to maximize the total number of restaurants. Among all optimal placements, find all pairs $(a, b)$ such that $a$ restaurants belong to Mac Donalds and $b$ to Burger Bing, with $a + b$ maximized.

Print all such pairs sorted by increasing $a$.

\subsection*{Input}

The first line contains a single integer $n~(3 \leq n \leq 2000)$.

Then follows $n-1$ lines: edges $x_i~y_i~(1 \leq x_i,~y_i \leq n)$.

The edges form a tree.

\subsection*{Output}

\begin{verbatim}
z
a1 b1
a2 b2
...
az bz
\end{verbatim}

Where $z$ is the count of pairs, and $(a_i, b_i)$ are all pairs with maximum $a_i+b_i$, sorted by ai ascending. If no valid placement uses both networks, print $0$ only.

\subsection*{Explanation of the new rules}

\begin{itemize}
    \item Adjacent restaurants cannot be different brands (so if both ends of an edge have restaurants, they must be the same brand).
    \item However, for any chain $u - v - w$, if $u$ and $w$ both have restaurants, then they must be different brands (distance-2 same-brand is forbidden).
    \item Together these imply strong local coupling between siblings in the rooted tree:
    \begin{itemize}
        \item If a node is occupied, then at most one of its children can also be occupied at its root (and it must have the same brand as the parent).
        \item If a node is empty, then among its children, at most one may have a red root and at most one may have a blue root (siblings with occupied roots must have different brands).
    \end{itemize}
\end{itemize}

These constraints make the original “remove one vertex and do subset sums” approach invalid; we need a new DP.

\subsection*{Examples}

\textbf{Input}
\begin{verbatim}
5
1 2
2 3
3 4
4 5
\end{verbatim}

\textbf{Output}
\begin{verbatim}
2
1 2
2 1
\end{verbatim}
One optimal solution uses 3 restaurants on the path (e.g., 1=R, 2=R, 4=., 5=B), giving totals $(2,1)$ or $(1, 2)$.

\textbf{Input}
\begin{verbatim}
5
1 2
1 3
1 4
1 5
\end{verbatim}

\textbf{Output}
\begin{verbatim}
1
1 1
\end{verbatim}
A star: the center must be empty to allow both brands on leaves; at most one red leaf and one blue leaf → exactly $(1,1)$.
\section{Near-Valid Test Case Example: Perpendicular Segments time }
\label{appendix:test_case_gen_case_study}

\textbf{time limit per test:} 2 seconds \\
\textbf{memory limit per test:} 256 megabytes

You are given a coordinate plane and three integers $X$, $Y$, and $K$. Find two line segments $AB$ and $CD$ such that
\begin{enumerate}
  \item the coordinates of points $A,B,C,$ and $D$ are integers;
  \item $0 \le A_x,B_x,C_x,D_x \le X$ and $0 \le A_y,B_y,C_y,D_y \le Y$;
  \item the length of segment $AB$ is at least $K$;
  \item the length of segment $CD$ is at least $K$;
  \item segments $AB$ and $CD$ are perpendicular: if you draw lines that contain $AB$ and $CD$, they will cross at a right angle.
\end{enumerate}

\noindent Note that it's \textbf{not} necessary for segments to intersect. Segments are perpendicular as long as the lines they induce are perpendicular.

\subsection*{Input}
The first line contains a single integer $t$ ($1 \le t \le 5000$) --- the number of test cases. Next, $t$ cases follow.

\noindent The first and only line of each test case contains three integers $X$, $Y$, and $K$ ($1 \le X,Y \le 1000;\; 1 \le K \le 1414$).

\noindent\textit{Additional constraint on the input:} the values of $X$, $Y$, and $K$ are chosen in such a way that the answer exists.

\subsection*{Output}
For each test case, print two lines. The first line should contain 4 integers $A_x, A_y, B_x,$ and $B_y$ --- the coordinates of the first segment.

\noindent The second line should also contain 4 integers $C_x, C_y, D_x,$ and $D_y$ --- the coordinates of the second segment.

\noindent If there are multiple answers, print any of them.

\subsection*{Example}
\paragraph{input}
\begin{verbatim}
4
1 1 1
3 4 1
4 3 3
3 4 4
\end{verbatim}

\paragraph{output}
\begin{verbatim}
0 0 1 0
0 0 0 1
2 4 2 2
0 1 1 1
0 0 1 3
1 2 4 1
0 1 3 4
0 3 3 0
\end{verbatim}

\subsection{NEAR-VALID TEST CASE EXAMPLE}

Here are two minimal test inputs that differ by only 1 in a single field ($K$). The first is eligible (a solution exists), the second is not (no segment of length $\geq K$ can fit at all).

\paragraph{Valid} 
\begin{verbatim}
1
4 4 5
\end{verbatim}
Why: In a 4×4 box, you can take two perpendicular diagonals: $(0,0)-(4,4)$ and $(0,4)-(4,0)$. Each has length $\sqrt{4^2 + 4^2} = \sqrt{32} \approx 5.657 \geq 5$.

\paragraph{Near Valid}
\begin{verbatim}
1
4 4 6
\end{verbatim}
Why: The longest possible segment inside a 4×4 box is the diagonal 5.657, which is $<$ 6, so even one segment of length $\geq$ 6 can’t exist, hence two perpendicular ones can’t either.
\section{Example of incorrect mutant programs: Guess The Tree}
\label{appendix:wrong_solution}

\textbf{time limit per test:} 2 seconds\\
\textbf{memory limit per test:} 256 megabytes

\noindent\textit{This is an interactive problem.}

Misuki has chosen a secret tree with $n$ nodes, indexed from $1$ to $n$, and asked you to guess it by using queries of the following type:
\begin{itemize}
  \item \verb|"? a b"| --- Misuki will tell you which node $x$ minimizes $\lvert d(a,x)-d(b,x)\rvert$, where $d(x,y)$ is the distance between nodes $x$ and $y$. If more than one such node exists, Misuki will tell you the one which minimizes $d(a,x)$.
\end{itemize}

Find out the structure of Misuki's secret tree using at most $15n$ queries!

\subsection*{Input}
Each test consists of multiple test cases. The first line contains a single integer $t$ ($1\le t\le 200$) --- the number of test cases.

\noindent Each test case consists of a single line with an integer $n$ ($2\le n\le 1000$), the number of nodes in the tree.

\noindent It is guaranteed that the sum of $n$ across all test cases does not exceed $1000$.

\subsection*{Interaction}
The interaction begins by reading the integer $n$.

\noindent Then you can make up to $15n$ queries.

\noindent To make a query, output a line in the format \verb|"? a b"| (without quotes) ($1\le a,b\le n$). After each query, read an integer --- the answer to your query.

\noindent To report the answer, output a line in the format \verb|"! a1 b1 a2 b2 ... a_{n-1} b_{n-1}"| (without quotes), meaning that there is an edge between nodes $a_i$ and $b_i$, for each $1\le i\le n-1$. You can print the edges in any order.

\noindent After $15n$ queries have been made, the response to any other query will be $-1$. Once you receive such a response, terminate the program to receive the \texttt{Wrong\_Answer} verdict.

After printing each line, do not forget to output the end of line and flush the output buffer. Otherwise, you will receive the \texttt{Idleness limit exceeded} verdict. To flush, use:
\begin{itemize}
  \item \verb|fflush(stdout)| or \verb|cout.flush()| in C++;
  \item \verb|System.out.flush()| in Java;
  \item \verb|flush(output)| in Pascal;
  \item \verb|stdout.flush()| in Python;
  \item see the documentation for other languages.
\end{itemize}

\subsection*{Hacks}
For hacks, use the following format: The first line contains an integer $t$ ($1\le t\le 200$) --- the number of test cases.

\noindent The first line of each test contains an integer $n$ --- the number of nodes in the hidden tree.

\noindent Then $n-1$ lines follow. The $i$-th of them contains two integers $a_i$ and $b_i$ ($1\le a_i, b_i\le n$), meaning that there is an edge between $a_i$ and $b_i$ in the hidden tree.

\noindent The sum of $n$ over all test cases must not exceed $1000$.

\subsection*{Example}
\paragraph{input}
\begin{verbatim}
1
4
1
1
3
\end{verbatim}

\paragraph{output}
\begin{verbatim}
? 1 2
1
? 1 3
1
? 1 4
3
! 1 2 1 3 3 4
\end{verbatim}

\subsection*{Note}
A tree is an undirected acyclic connected graph. A tree with $n$ nodes will always have $n-1$ edges.

\noindent In the example case, the answer to \verb|"? 1 2"| is $1$. This means that there is an edge between nodes $1$ and $2$.

\noindent The answer to \verb|"? 1 3"| is $1$. This means that there is an edge between nodes $1$ and $3$.

\noindent The answer to \verb|"? 1 4"| is $3$. It can be proven that this can only happen if node $3$ is connected to both node $1$ and $4$.

\noindent The edges of the tree are hence $(1,2)$, $(1,3)$ and $(3,4)$.

\subsection{Why it's Wrong}

Here’s a subtly broken version. It behaves the same on the sample, but the component update decision is based on comparing the node index to the label of the found node (instead of comparing component labels). This will make it fail on plenty of hidden cases while still looking plausible.

\begin{lstlisting}[style=cpp]
#include <iostream>
#include <numeric>
#include <vector>
#include <array>

using namespace std;

int query(int u, int v) {
  cout << "? " << u + 1 << ' ' << v + 1 << '\n';
  int x; cin >> x;
  return x - 1;
}

int main() {
  int t; cin >> t;
  while (t--) {
    int n; cin >> n;

    vector<array<int, 2>> e;
    vector<int> c(n);
    iota(c.begin(), c.end(), 0);

    auto addEdge = [&](int u, int v) {
      e.push_back({u, v});
      vector<int> cand;
      for (int i = 0; i < n; i++)
        if (c[i] == c[v])
          cand.emplace_back(i);
      for (int i : cand)
        c[i] = c[u];
    };

    for (int i = 0; i < n - 1; i++) {
      int u = 0, v = 0;
      while (c[v] == c[u]) v++;
      int x;
      // SUBTLE BUG: use (u == c[x]) instead of (c[u] == c[x]).
      while ((x = query(u, v)) != u) {
        if (u == c[x]) u = x;   // <-- wrong criterion
        else v = x;
      }
      addEdge(u, v);
    }

    cout << "!";
    for (auto [u, v] : e) cout << ' ' << u + 1 << ' ' << v + 1;
    cout << '\n';
  }
}
\end{lstlisting}
\section{Human Expert Grading Policy}
\label{appendix:grade}
Each problem is assessed independently across five dimensions: \text{Solvability}, \text{Solution Correctness}, \text{Quality Rating}, \text{Novelty}, and \text{Difficulty}.

\subsection{Solvability}
This criterion evaluates whether the problem admits a deterministic solution that meets the given time and space constraints.
\begin{itemize}
    \item \textbf{Yes:} The problem has a deterministic solution within the prescribed complexity limits.
    \item \textbf{No:} It can be proven that no deterministic solution exists within the constraints.
    \item \textbf{Unknown:} It is currently indeterminate whether a valid deterministic solution exists.
\end{itemize}

\subsection{Solution Correctness}
This evaluates whether the provided reference solution (e.g., a model-generated code) is fundamentally correct in its approach.
\begin{itemize}
    \item \textbf{Yes:} The core idea of the solution is correct. Minor implementation details may contain errors but do not invalidate the overall method.
    \item \textbf{No:} The solution contains fundamental flaws in reasoning or design.
    \item \textbf{Unknown:} It is not possible to determine the correctness of the provided solution approach.
\end{itemize}

\subsection{Quality Rating}
The overall quality of a problem is rated on a scale from 1 to 5, considering clarity, originality of approach, and implementation difficulty.
\begin{itemize}
    \item \textbf{1:} Severely flawed. For example, unsolvable problems or incomprehensible descriptions.
    \item \textbf{2:} Low quality. For example, unclear statements, trivial or uninteresting solutions, or unverifiable correctness of the reference solution.
    \item \textbf{3:} Moderate quality. The problem is clear and solvable, but may feel derivative, overly difficult for humans without conceptual elegance, or overly focused on obscure knowledge.
    \item \textbf{4:} High quality. The problem is clear, solvable, moderately difficult, and contains at least one clever or insightful idea, though it may still exhibit some formulaic aspects.
    \item \textbf{5:} Excellent quality. The problem is competition-grade: elegant, novel, appropriately challenging, and free from contrived traps or rote patterns.
\end{itemize}

\subsection{Novelty}
This measures whether the problem introduces a genuinely new challenge.
\begin{itemize}
    \item \textbf{Yes:} The problem is novel and not seen in prior contests or problem archives, even for experienced setters.
    \item \textbf{No:} The problem duplicates or closely resembles existing ones.
    \item \textbf{Unknown:} Uncertainty remains regarding the existence of prior similar problems.
\end{itemize}

\subsection{Difficulty}
Difficulty is assessed using the Codeforces rating scale (e.g., 800, 1200, 1600, \dots).  
This score is provided as a reference and does not directly influence the overall rating.

\subsection{Overall Problem Rating}
A composite grading system is defined as follows. A problem must satisfy the requirements of its assigned level and all lower levels.

\begin{itemize}
    \item[] $\geq$ \textbf{0:} All evaluated problems.
    \item[] $\geq$ \textbf{1:} Solvability = Yes.
    \item[] $\geq$ \textbf{2:} Solution Correctness = Yes.
    \item[] $\geq$ \textbf{3:} Quality Rating $\geq 3$.
    \item[] $\geq$ \textbf{4:} Quality Rating $\geq 4$.
    \item[] $\geq$ \textbf{5:} Novelty = Yes.
    \item[] $\geq$ \textbf{6:} Quality Rating $\geq 5$.
\end{itemize}
\section{Simulator in Interactor}
\label{appendix:interactor}

\begin{algorithm}[H]
\caption{\textsc{Simulate} (in Interactor)}
\KwInput{Interactor $I$, solver $\mathcal{A}$, Validator $V^\star$}
\KwOutput{\textsf{Accepted}/\textsf{Rejected}}
Initialize protocol state; set time/memory limits; seed RNG deterministically\;
\While{session active}{
  $msg_I \leftarrow I(\text{state})$; \If{$\neg$ \textsc{WellFormed}$(msg_I)$}{\Return \textsf{Rejected}}
  $msg_A \leftarrow \mathcal{A}(msg_I)$; \If{$\neg$ \textsc{WellFormed}$(msg_A)$}{\Return \textsf{Rejected}}
  \If{\textsc{ProvidesFinalAnswer}$(msg_A)$}{
     \If{$V^\star(\text{implied input}) \neq \textsf{valid}$}{\Return \textsf{Rejected}}
     \Return \textsc{JudgeFinal}$(I, msg_A)$
  }
  \textsc{UpdateState}$(\text{state}, msg_I, msg_A)$\;
}
\Return \textsf{Rejected}\;
\end{algorithm}
\section{Limitations and Future Work}
\label{appendix:limitation}


While \model demonstrates a robust pipeline for generating and verifying competitive programming problems, we acknowledge several limitations that open up promising avenues for future research.

\paragraph{The quality judgment gap.} A primary limitation is that LLMs currently lack a robust, human-aligned sense of problem quality. As shown in Finding 4 of Section~\ref{sec:problem_generation}, there is little correlation between LLM self-evaluations and the scores provided by human experts. Consequently, our system still requires human-in-the-loop to distinguish truly high-quality and novel problems from those that are merely technically correct. This reliance on manual annotation is a bottleneck to a scalable high-quality problem generation pipeline. A key future direction is to develop a specialized judge LLM, fine-tuned on expert-rated problem datasets to better align its assessments with human preferences, potentially serving as a reliable automated filter.

\paragraph{The bottleneck for automated self-improvement.} \model is ideally suited for creating a self-improvement loop, where an LLM could enhance its reasoning abilities by training on the problems it generates. However, deploying and testing such a system faces immediate practical barriers. The most capable problem-generating models like \texttt{GPT-5} are closed-source and do not expose the APIs necessary for reinforcement learning. Even for powerful open-source models, scalable and stable RL frameworks are not yet mature to support fine-tuning at this scale to the best of our knowledge. A natural future direction is to focus on developing RL or self-play frameworks compatible with frontier LLMs to unlock the full potential of \model.

\paragraph{Advancing beyond recombinational novelty.} Our analysis indicates that LLMs currently tend to create new problems by combining existing algorithms rather than creating conceptually new problems that require subtle insights. Future research could explore novel prompting strategies or problem-generation pipelines to encourage greater conceptual leaps. Additionally, for valid problems that LLMs can generate but cannot solve themselves (i.e., their reference solution std.cpp is incorrect), \model's strict verification protocol currently filters them out. A prominent future direction is to develop scalable methods for identifying these valid but unsolved questions without relying on manual check by human experts, which is an extremely labor-intensive process.


\end{document}